# Non-equilibrium Thermal Super-radiation of Real Materials

Yiling Yu[2], Linyou Cao[1,2]*

[1]Department of Materials Science and Engineering, North Carolina State University, Raleigh NC 27695;
[2]Department of Physics, North Carolina State University, Raleigh NC 27695;

**Abstract:**

We elucidate the theoretically maximal thermal radiation power from real materials at a given temperature. Our results demonstrate that the thermal radiation from real materials may be larger than the blackbody emission in free space, and indicate that this is rooted in the high refractive index of the materials. The refractive index contrast between the materials and environment dictates the radiation of real materials genetically not under thermodynamic equilibrium, but on the other hand can give rise to a larger density of photonic modes than that of the blackbody. One key to maximize the thermal radiation is to minimize the impedance mismatch of the materials with environment. By following this principle, we present a design of a carbon core coated by a four-layer transparent shell with gradually changed refractive indexes that can emit > 30 times more power than the blackbody, which reasonably approaches the predicted radiation maximum.

* To whom correspondence should be addressed.

Email: lcao2@ncsu.edu

Maximizing the radiation power from a thermal emitter promises the ultimate improvement in the efficiency of incandescence lighting[1], radiative cooling or heating [2-5], thermal infrared sources [6-8], and thermophotovoltaics [9-13]. It is generally believed that no emitter made of real materials can radiate more power than a blackbody. The blackbody emission essentially defines the thermal radiation under thermodynamic equilibrium. In stark contrast with this conventional wisdom, here we demonstrate that the thermal radiation of real materials, such as carbon, can exceed the blackbody emission in free space by more than one order of magnitude. The radiation from real materials is genetically non-equilibrium due to the refractive index contrast between the materials and environment.

The difference between the blackbody emission and the non-equilibrium radiation of real materials can be best understood from a perspective of mode coupling. Without losing generality, we examine the thermal radiation of a spherical emitter in free space (Fig. 1). Similar to the strategy of previous studies [14], we consider the thermal radiation as an out-coupling of the electromagnetic energy stored in the photonic modes of the emitter to the surrounding medium. For the convenience of discussion, we refer the photonic modes in the emitter as emitter modes. The power $P_j$ radiated from an arbitrary emitter mode $j$ can be intuitively correlated with the energy $E$ stored at the mode (modal energy) as $P_j \propto \gamma E$. $\gamma$ is the radiative decay rate of the mode and physically indicates the coupling efficiency between the emitter mode and the medium. The total radiation power of the emitter is a simple sum of the contribution from all the emitter modes as $P_{total} = \sum_j P_j = \int \rho(\lambda)\gamma E d\lambda$, where $\rho(\lambda)$ is the density of emitter modes at the wavelength $\lambda$. Three parameters dictate the process of thermal radiation, the modal energy $E$, the radiative decay rate $\gamma$, and the density of photonic modes in the emitter $\rho(\lambda)$. The real emitter differentiates with the blackbody in all the three parameters.

The blackbody can be considered as an ideal photon gas with a uniform spatial distribution of photonic modes (Fig. 1a). The mode density in the blackbody emitter is identical to that of the medium, $\rho_b = 8\pi V_b/\lambda^4$, where $V_b$ is the volume of the emitter. Under thermodynamic equilibrium the energy of photons follows a Maxwell-Boltzman distribution. As a result, the energy stored at each mode of the blackbody emitter is $E_b = \hbar\omega/[(e^{\hbar\omega/kT}-1)]$, where $\hbar$ is the Planck constant, $k$ is the Boltzman constant, $T$ is the temperature, and $\omega$ is the angular frequency. By matching with the well-established expression for blackbody emission $\rho_b\gamma_b E_b = 2\pi S_b c/\lambda^4 \cdot \hbar\omega/[(e^{\hbar\omega/kT}-1)]$, we can find the radiative decay $\gamma_b = c/4 \cdot S_b/V_b = 3c/4r$, where $r$ and $S_b$ are the radius and surface area of the blackbody emitter. In brief, the modal energy, radiative decay rate, and mode density of the spherical blackbody emitter are $\hbar\omega/[(e^{\hbar\omega/kT}-1)]$, $3c/4r$, and $8\pi V_b/\lambda^4$, respectively.

We can find out the parameters for real emitters using a model that we have recently developed, coupled leaky mode theory (CLMT). The CLMT model was originally developed for the evaluation of light absorption [15-17], but can be used for the analysis of thermal radiation as well. As stated by the Kirchhoff's law, the emissivity of an object is identical to the absorption efficiency of the object [18]. The CLMT model suggests that the modes involved in the radiation of real emitters are leaky modes, which are photonic modes with propagating waves outside the emitter. The density of leaky modes in a spherical emitter can be reasonably approximated as $\rho_R(\lambda) = n^3 8\pi V_R/\lambda^4$, where $V_R$ and $n$ are the volume and refractive index of the emitter, respectively [19]. Each of the leaky modes is featured with a complex eigenvalue $(N_{real} - N_{imag} \cdot i)$ that can be solved analytically or numerically [15-17]. The radiative decay rate of the leaky mode is related with the imaginary part of the eigenvalue as $\gamma_R = c \cdot N_{imag}/(n \cdot r)$ [15]. Intuitively, the radiative decay rate in the blackbody imposes an upper limit for $\gamma_R$, $2\gamma_R < \gamma_b$, in which the constant of 2 arises from the expression of the radation power defined by the CLMT model from real materials as $P_R = 2\gamma_R E_R$ [15]. This constrain on the radiative decay rate of leaky modes is slightly different from what was derived in a previous study [20]. The smaller radiative decay in real emitters is due to the impedance mismatch between the emitter and its medium, which confines the radiation inside with only part of the energy being coupled outward (Fig. 1b).

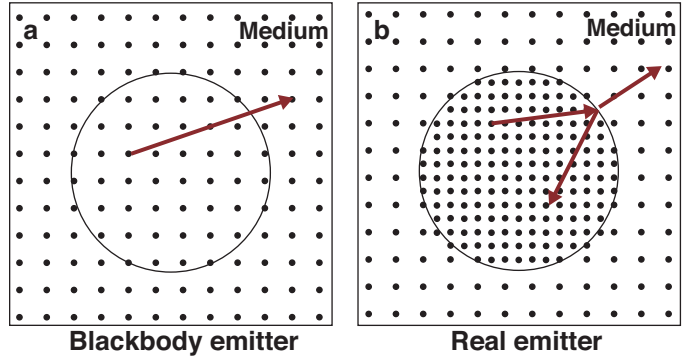

Figure 1. Schematic illustration of the mode coupling perspective for thermal radiation. (a) a blackbody emitter and (b) a real emitter in free space. The circle indicates the boundary between the emitter and the medium, and the dots represent photonic modes with the interspacing indicating the density of the modes. The red arrow illustrates the out-coupling of the modal energy from the emitter to the medium.

The energy stored at each leaky mode in the real emitter $E_R$ can be derived by using the CLMT model as (see Supporting Materials for the detailed derivative process)

$$E_R = \frac{1}{\pi} \int_0^\infty \frac{\gamma_{abs}(\omega)}{(\omega-\omega_0)^2 + (\gamma_R + \gamma_{abs}(\omega))^2} \cdot Corr \cdot \frac{\hbar\omega}{e^{\hbar\omega/kT}-1} d\omega \quad (1)$$

where $\omega_0$ is the eigen frequency of the mode, and $\gamma_{abs}(\omega)$ is the decay caused by the intrinsic absorption of the materials at the frequency $\omega$, $Corr$ is a correction term for accurately evaluating the emissivity of real materials. There is no rigorous expression for the $Corr$, but we have previously demonstrated that $1/[1+4(\omega/\omega_0-1)^2]$ for $\omega > \omega_0$ and $1/[1+4(\omega_0/\omega-1)^2]$ for $\omega < \omega_0$ is a reasonable approximation[17]. Unlike the modal energy of the black body emitter $E_b$, which is only related with the eigen frequency of the mode, $E_R$ has a format of integral over frequencies. This is because the leaky mode of real emitters can radiate a range of frequencies due to a finite photon confinement. For materials with infinitesimal intrinsic absorption and radiative decay, $\omega \gg \gamma_{abs} \gg \gamma_R$, the integrand of eq.(1) is $\hbar\omega_0/[(e^{\hbar\omega_0/kT}-1)]$, identical to the modal energy of the blackbody emitter. However, for typical emitting materials with non-trivial absorption and radiative decay, the modal energy $E_R$ may substantially deviate from the modal energy in blackbody $E_b$. To illustrate this notion, we numerically evaluate $E_R$ as functions of $\gamma_{abs}$ and $\gamma_R$ with the resonant frequency $\omega_0$ and the temperature $T$ arbitrarily set to be $3.14 \times 10^{14}$ Hz and 350 K (Fig. 2). We can find that $E_R$ is generally smaller than $E_b$, in particular, at large $\gamma_R$ and small $\gamma_{abs}$ (Fig. 2b). This can be intuitively understood from the perspective of energy conservation. The energy stored at the leaky mode tends to be small when the mode radiates energy strongly (large $\gamma_R$) but absorbs energy weakly (small $\gamma_{abs}$). The smaller modal energy than what is dictated by the Maxwell-Boltzmann distribution indicates that the thermal radiation of real emitters is not under thermodynamic equilibrium!

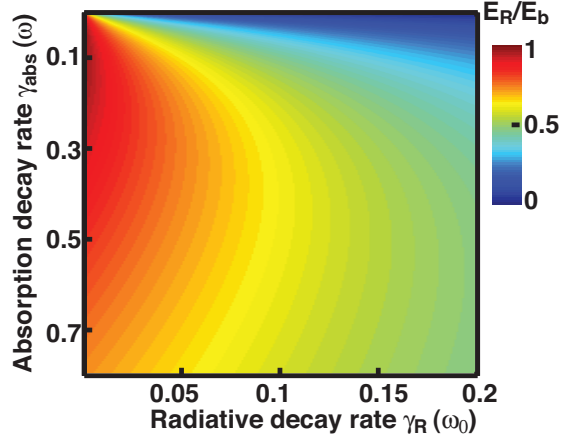

Figure 2. The modal energy $E_R$ in real emitters. The modal energy is normalized to the modal energy in the blackbody $E_b = \hbar\omega_0/[(e^{\hbar\omega_0/kT}-1)]$. It is plotted as a function of the absorption decay rate $\gamma_{abs}$ (the vertical axis) and the radiative decay rate $\gamma_R$ (the horizontal axis). The resonant frequency $\omega_0$ and the temperature $T$ are arbitrarily set to be $3.14 \times 10^{14}$ Hz and 350 K. The absorption decay rate is in unit of the emission wavelength $\omega$ and the radiative decay rate is in unit of the resonant wavelength $\omega_0$.

With the knowledge of the three parameters in both real and blackbody emitters, we can find that the radiation from the real emitter can in principle be larger than the blackbody emission. For the convenience of comparison, we re-write the expressions for the radiation from the real emitter and blackbody as $P_R = \int \rho_R (2\gamma_R) E_R d\lambda$ and $P_b = \int \rho_b \gamma_b E_b d\lambda$, respectively. While the real emitter has a smaller radiative decay $\gamma_R$ and modal energy $E_R$ at each mode, its mode density $\rho_R$ may be much larger than that of the blackbody emitter due to the high refractive index of typical emitter materials. Therefore, if the gain in the mode density could offset the loss caused by the smaller radiative decay and modal energy, the real emitter would be able to radiate more power than the blackbody emitter.

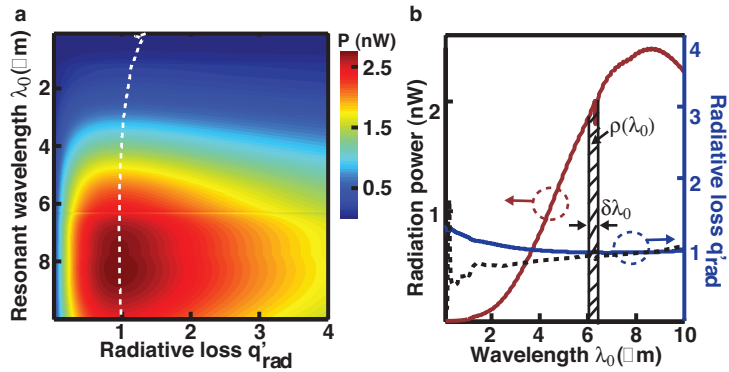

Figure 3. Thermal radiation of a single-mode carbon emitter. (a) The thermal radiation power as a function of the radiative loss (the horizontal axis) and the resonant wavelength (the vertical axis). The white dashed line indicates the optimal radiation at each resonant wavelength. (b) The optimal single-mode radiation (red line) and associated radiative loss (blue line) as a function of the resonant wavelength. The intrinsic absorption loss of carbon materials is also given (dashed black line). The shaded area is to schematically illustrate the integration of the contributions from multiple leaky modes.

As a further support, we quantitatively evaluate the maximal thermal radiation from a real emitter. We start with examining the thermal radiation from a single-mode emitter, which can be written as

$$P_R = \gamma_R E_R = -\frac{1}{\pi}\int_0^\infty \frac{q'_{rad}\lambda_0\gamma_{abs}(\lambda)}{\left(\frac{\lambda_0}{\lambda}-1\right)^2 + \left(q'_{rad}+\frac{\lambda_0\gamma_{abs}(\lambda)}{2\pi c}\right)^2} \cdot Corr \cdot \frac{hc/\lambda^3}{e^{hc/\lambda kT}-1}d\lambda \quad (2)$$

In eq. (2) the integral is changed to be over wavelength instead of frequency as in eq. (1), and a new parameter $q'_{rad}$ is introduced as $q'_{rad} = \gamma_R/\omega_0$. These changes are simply for the convenience of discussion and numerical evaluation. $q'_{rad}$ is also equal to the ratio of the imaginary ($N_{imag}$) and real ($N_{real}$) parts of the eigenvalue, $q'_{rad} = \gamma_R/\omega_0 = N_{imag}/N_{real}$ [15]. We refer it as the radiative loss of leaky modes. For a single-mode emitter with known intrinsic absorption $\gamma_{abs}$ and temperature $T$, the thermal radiation is only dependent on two variables, the resonant frequency $\lambda_0$ and the radiative loss $q'_{rad}$ of the mode. We use a carbon emitter with a temperature of 350 K as an example to illustrate this notion. Fig. 3a shows the calculated radiation power of the single-mode carbon emitter as functions of $q'_{rad}$ and $\lambda_0$.

The maximal thermal radiation (super-radiation) from typical emitters, which involve multiple modes, can be derived from the calculation result of the single-mode emitter. The multiple modes typically have different resonant wavelengths. To maximize the thermal radiation in a multi-mode emitter would request the radiation of each mode to be optimized. From Fig. 3a we can find out the optimal radiation power of a single leaky mode with an arbitrary resonant wavelength, as indicated by the white dashed line. The single-mode optimal radiation is replotted as a function of the resonant wavelength in Fig. 3b. Also plotted is the radiative loss associated with the optimal radiation, which can be found reasonably matching the intrinsic absorption loss $n_{imag}/n_{real}$ ($n_{real}$, $n_{imag}$ are the real and imaginary part of the refractive index, respectively) of carbon materials (Fig. 3b) [21]. This match is analogue to the critical coupling in waveguide-resonator systems that is known able to enable efficient electromagnetic coupling [22].

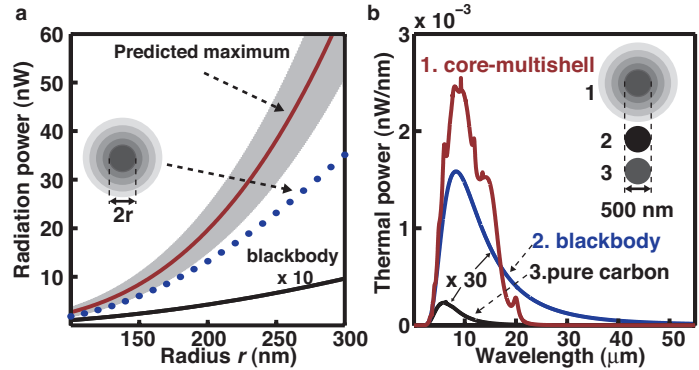

Figure 4. Super-radiation of real emitters. (a) The predicted maximal radiation of a spherical carbon emitter at a temperature of 350 K as a function of the radius (red line). An estimated 20% error is included as indicated by the shaded area. The blackbody emission with the same temperature is also given but the result is multiplied by 10 for visual convenience (black). The dots indicate the thermal emission from a carbon core coated by four transparent shells as a function of the radius of the carbon core. (b) Spectral thermal radiation of the core-multishell structure (structure 1), the blackbody (structure 2) and a carbon particle without coating (structures 3). The radius of the emitting part of these structures are all set to be 250 nm. The results for the structure 2 and 3 are multiplied by 30 for visual convenience.

We can obtain the maximal thermal radiation of multi-mode emitters by integrating over the single-mode optimal radiation as $P_R = \int P_{opt}(\lambda_0)\rho(\lambda_0)d\lambda_0$, where $P_{opt}(\lambda_0)$ and $\rho(\lambda_0)$ are the single-mode optimal radiation and the density of leaky modes at an arbitrary resonant wavelength $\lambda_0$. Fig. 4a shows the maximal radiation of a spherical carbon emitter as a function of the radius, which is calculated using the result in Fig. 3b. Note that the constrain for the radiative decay in real emitters $2\gamma_R < \gamma_b$ imposes an upper limit on $q'_{rad}$ as $q'_{rad} < 3\lambda_0/(16\pi r)$. For large $r$ or small wavelength $\lambda_0$ where the $q'_{rad}$ associated with the optimal radiation shown in Fig. 3b may be larger than this upper limit, we count the radiation power associated with the maximal radiative loss of $3\lambda_0/(16\pi r)$. An estimated error of 20% is included in the calculation results. It mainly originates from the imperfect accuracy of the CLMT model and a finite integration over the resonant wavelength. The integration would ideally be over the entire wavelength range rom 0 to infinity, but in reality the numerical integration is always performed in a finite wavelength range. The result in Fig.4a is obtained by integrating from 0.1 μm to 55 μm, in which the refractive index of the emitter materials is available [21]. The blackbody remission in free space with the same surface area and temperature ($\sigma T^4 \cdot 4\pi r^2$, $\sigma$ is the Stefan-Boltzman constant) is also plotted in Fig. 4a as a reference. Very significantly, the maximal radiation of the carbon emitter is far larger than the blackbody emission by more than one order of magnitude!

We can design emitters whose radiation may reasonably approach the predicted maximum. The result in Fig.3b indicates that to maximize the thermal radiation in real materials requests the radiative loss of leaky modes to be tuned reasonably matching the intrinsic absorption loss of the materials. However, the radiative loss of the leaky modes in typical structures is much smaller than the intrinsic absorption loss of typical emitting materials [15, 19], which is due to the impedance mismatch between the emitter and environment that can provide efficient photon confinement. The low radiative loss of leaky modes is one major reason why the thermal radiation from real materials is often much less than the blackbody emission. We have previously demonstrated that coating emitting materials with a transparent shell in gradually-changed refractive index presents a promising strategy to increase the radiative loss of leaky modes [19, 23]. The gradually changed refractive index can substantially mitigate the impedance mismatch to facilitate the radiation. Without extensive optimization, we design a core-multishell spherical structure that consists of a carbon core coated by four layers of transparent materials. The thickness of the four shells from inner to outer are set to be 1425 nm, 400 nm, 300 nm, and 900 nm, and the refractive index 5, 3.35, 2.4, and 1.75, respectively. We use the well-established Mie theory to evaluate the radiation power of the structure [24], and plat the result as a function of the radius of the carbon core (dots) in Fig. 4a. We can find that the thermal radiation of the designed structure is indeed larger than the blackbody emission by more than one order of magnitude (> 30 times). The dramatic enhancement in the thermal radiation can be more clearly seen in Fig. 4b, which show the radiation spectra of three different structures: the core-multishell structure, the blackbody, and a pure carbon particle. The radius of the emitting part in all these structures is set to be 250 nm.

The superior thermal radiation of the designed structure, although still smaller than the predicted maximum, strongly supports our analysis of the thermal super-radiation of real materials. The blackbody emission in free space is not the upper limit for the thermal radiation of real emitters. Our design also confirms that key to maximize thermal radiation is to mitigate the impedance mismatch of the emitter to boost the radiative loss of leaky modes. This principle is also confirmed by a recent study that demonstrates thermal radiation beyond the blackbody emission by covering a carbon emitter with a transparent dome [25]. It is worthwhile to point out that the given design is by no means the best one. We believe that optimizing the refractive profile in the transparent shell may further improve the thermal radiation.

*Supporting Materials*
*for*

# Non-equilibrium Thermal Super-radiation of Real Materials

Yiling Yu[2], Linyou Cao[1,2]*

[1]Department of Materials Science and Engineering, North Carolina State University, Raleigh NC 27695; [2]Department of Physics, North Carolina State University, Raleigh NC 27695;

* To whom correspondence should be addressed.
Email: lcao2@ncsu.edu

**This PDF file includes**

Derivative for the modal energy of leaky modes

References

**Derivative for the modal energy of leaky modes**

The thermal radiation $P_R$ of a spherical particle can be calculated from the absorption efficiency $Q_{abs}$ of the particle as [S1]

$$P_R = \int_0^\infty Q_{abs}(\omega) \frac{\omega^2}{4\pi^2 c^2} \frac{\hbar\omega}{e^{\hbar\omega/kT}-1} \cdot S \, d\omega \quad (1)$$

where $\omega$ is the emitting frequency, $c$ is the speed of light, $k$ is the Boltzman constant, $T$ is the temperature, and $S$ is the surface area of the particle and related with its radius as $S = 4\pi r^2$. The absorption efficiency $Q_{abs}$ of an arbitrary leaky mode $j$ in the spherical particle can be evaluated using the CLMT model as

$$Q_{abs,j} = \frac{2\gamma_{abs}\gamma_R}{(\omega-\omega_0)^2 + (\gamma_{abs}+\gamma_R)^2} \frac{c^2}{(\omega r)^2} \cdot Corr \quad (2)$$

where $\omega_0$ and $\gamma_R$ is the eigen frequency and the radiative decay rate of the mode, $\gamma_{abs}$ is the intrinsic absorption decay of the materials at the frequency $\omega$, and $Corr$ is the correction term as discussed in the main text. Typically, the particle would involve multiple leaky modes. In this case, the total absorption efficiency $Q_{abs}$ is a simple sum of the absorption efficiency contributed by each individual mode, and can be evaluated using the CLMT model as

$$Q_{abs} = \int_0^\infty Q_{abs,j}\rho(\omega_0)d\omega_0 = \int_0^\infty \frac{2\gamma_{abs}\gamma_R}{(\omega-\omega_0)^2 + (\gamma_{abs}+\gamma_R)^2} \frac{c^2}{(\omega r)^2} \cdot Corr \cdot \frac{\omega_0^2 n^3}{\pi^2 c^3} V d\omega_0 \quad (3)$$

By substituting eq. (3) in eq. (1), we may have

$$P_{total} = \int_0^\infty \int_0^\infty \frac{2\gamma_{abs}\gamma_R}{(\omega-\omega_0)^2 + (\gamma_{abs}+\gamma_R)^2} \frac{c^2}{(\omega r)^2} \cdot Corr \cdot \frac{\omega_0^2 n^3}{\pi^2 c^3} V \frac{\omega^2}{4\pi^2 c^2} \frac{\hbar\omega}{e^{\hbar\omega/kT}-1} \cdot S \, d\omega \, d\omega_0$$

$$= \int_0^\infty 2\gamma_R \left[ \frac{1}{\pi} \int_0^\infty \frac{\gamma_{abs}}{(\omega-\omega_0)^2 + (\gamma_{abs}+\gamma_R)^2} \cdot Corr \frac{\hbar\omega}{e^{\hbar\omega/kT}-1} d\omega \right] \frac{\omega_0^2 n^3}{\pi^2 c^3} V d\omega_0 \quad (4)$$

According to the CLMT model [S2], we set the radiation power by each leaky mode $P_R$ be correlated with the modal energy $E_R$ as $P_R = 2\gamma_R E_R$. The total radiation power is a simple sum of the contribution from all the leaky modes as

$$P_{total} = \int_0^\infty 2\gamma_R E_R \rho(\omega_0) d\omega_0 = \int_0^\infty 2\gamma_R E_R \frac{\omega_0^2 n^3}{\pi^2 c^3} V d\omega_0 \quad (5)$$

By comparing eq. (4) and (5), we can have

$$E_R = \frac{1}{\pi} \int_0^\infty \frac{\gamma_{abs}}{(\omega-\omega_0)^2 + (\gamma_{abs}+\gamma_R)^2} \cdot Corr \frac{\hbar\omega}{e^{\hbar\omega/kT}-1} d\omega \quad (6)$$

References

[S1]  G. W. Kattawar and M. Eisner, Appl. Optics 9 (1970).
[S2]  Y. Yu and L. Cao, Opt. Express 21 (2013).